\documentclass[fleqn,usenatbib]{mnras}

\usepackage{newtxtext,newtxmath}

\usepackage[T1]{fontenc}
\usepackage{ae,aecompl}

\usepackage{graphicx}	% Including figure files
\usepackage{amsmath}	% Advanced maths commands
\usepackage{amssymb}	% Extra maths symbols
\usepackage{hyperref}

\def\e{\boldsymbol{e}}

\def\k{\boldsymbol{k}}

\def\bnu{\boldsymbol{\nu}}
\def\btheta{\boldsymbol{\theta}}

\def\S{\boldsymbol{S}}

\title[Improving 21\,cm antenna calibration ]
{On the use of temporal filtering for mitigating galactic synchrotron calibration bias in 21 cm reionization observations}

\author[Charles et al.]
{
\parbox{\textwidth}{
Ntsikelelo Charles$^{1}$\thanks{g15c0896@campus.ru.ac.za}, Nicholas Kern$^{2}$, Gianni Bernardi$^{3,1,4}$, Landman Bester$^{4,1}$, Oleg Smirnov$^{1,4}$, Nicolas Fagnoni$^{5}$, Eloy de Lera Acedo$^{5,6}$}
\vspace{0.4cm} \\
\parbox{\textwidth}{
$^{1}$ Department of Physics and Electronics, Rhodes University, PO Box 94, Makhanda, 6140, South Africa\\
$^{2}$ Dept. of Physics and Kavli Institute for Astrophysics and Space Research, Massachusetts Institute of Technology, Cambridge, MA, USA\\
$^{3}$INAF-Istituto di Radio Astronomia, via Gobetti 101, 40129 Bologna, Italy\\
$^{4}$South African Radio Astronomy Observatory, Black River Park, 2 Fir Street, Observatory, Cape Town, 7925, South Africa\\
$^{5}$Cavendish Astrophysics, University of Cambridge, Cambridge CB3 0HE, UK\\
$^{6}$Kavli Institute for Cosmology in Cambridge, University of Cambridge, Cambridge, UK\\
}
}

\date{Accepted XXX. Received YYY; in original form ZZZ}

\pubyear{2022}

\begin{document}
\label{firstpage}
\pagerange{\pageref{firstpage}--\pageref{lastpage}}
\maketitle

\begin{abstract}

Precision antenna calibration is required for mitigating the impact of foreground contamination in 21\,cm cosmological radio surveys.
One widely studied source of error is the effect of missing point sources in the calibration sky model; however, poorly understood diffuse galactic emission also creates a calibration bias that can complicate the clean separation of foregrounds from the 21\,cm signal.
In this work, we present a technique for suppressing this bias with temporal filtering of radio interferometric visibilities observed in a drift-scan mode. We demonstrate this technique on mock simulations of the Hydrogen Epoch of Reionization Array (HERA) experiment. Inspecting the recovered calibration solutions, we find that our technique reduces spurious errors by over an order of magnitude. This improved accuracy approaches the required accuracy needed to make a fiducial detection of the 21\,cm signal with HERA, but is dependent on a number of external factors that we discuss. We also explore different types of temporal filtering techniques and discuss their relative performance and tradeoffs.

\end{abstract}


%%%%%%%%%%%%%%%%%%%%%%%%%%%%%%%%%%%%%%%%%%%%%%%%%%

%%%%%%%%%%%%%%%%% BODY OF PAPER %%%%%%%%%%%%%%%%%%

\section{Introduction}
\label{sec:intro}
The Hydrogen Epoch of Reionization Array (HERA, \citet{DeBoer2017}) is a newly built low-frequency radio interferometric array primarily designed to probe the Epoch of Reionization (EoR). The EoR period is one of the very difficult areas in cosmology to probe, and as of this writing, it is one of the least known areas experimentally. There are  several other low-frequency arrays aimed at probing the EoR; these include the Low-Frequency Array (LOFAR, \citet{VanHaarlem2013}), Murchison Widefield Array (MWA, \citet{Tingay2013}), Giant Metrewave Radio Telescope (GMRT) EoR experiment \citep{Paciga2013}, and the Square Kilometer Array \citep[SKA,][]{Koopmans2015}. 

The 21~cm emission from the hyperfine transition of neutral hydrogen is one of the best observational probes of the EoR period.  EoR experiments all face the common problems of bright foregrounds from Galaxy and extragalactic sources, which can be up to five orders of magnitude brighter than the 21~cm emission \citep[e.g.,][]{Jelic2008, Bernardi2010, Thyagarajan2016}. Fortunately, foreground emission is spectrally smooth compared to 21~cm emission, and therefore the 21~cm signal can be separated from foreground emission, for example, using the \textit{delay spectrum} approach \citep{Parsons2012a, Thyagarajan2013, Liu2014a}.
The delay spectrum uses interferometric delays to isolate the power spectrum of the 21~cm signal from the foregrounds. Due to the spectral nature of the 21~cm signal, its power spectrum has emission at all Fourier $k$ wavevector modes, whereas the foreground power spectra is limited to a wedge-like region in a two-dimensional $k_\parallel-k_\perp$-space, referred to as the foreground wedge \citep{Datta2010,Parsons2012a,Trott2012,Vedantham2012,Thyagarajan2013,Liu2014a,Liu2014b,Morales2019,Mevius2022}.
This separation, in principle, leads to a region of data at high $k_\parallel$ where the EoR 21\,cm signal should cleanly separate from foreground contamination, what is referred to as the ``EoR window.''

However, before power spectrum analysis is performed, interferometric data require calibration in order to correct for instrumental effects \citep[see, for example,][for a review on interferometric calibration]{Smirnov2011}. Calibration often includes the use of sky models, and typically, sky models are built from source catalogues. Source catalogues \citep[e.g.][]{Hurley-Walker2017, Shimwell2017, Hurley-Walker2022}, however, do not fully characterise the sky emission due to the inevitable limited sensitivity and angular resolution \citep{Grobler2014, Wijnholds2016, Trott2016, Procopio2017, Barry2021, Gehlot2021}. In addition, sky models often exclude diffuse Galactic synchrotron emission and 
are, therefore, not a full accounting of the low-frequency sky brightness distribution. This ``incompleteness'' in our understanding of the true sky leads to errors when performing calibration, which can compromise our ability to invert the telescope response and prohibit a clean measurement of the 21\,cm cosmological signal \citep{Barry2016, Patil2016, Ewall-Wice2017, Morales2019, Byrne2019, Kern2020a, Dillon2020, Byrne2021}. 

The instrumental response of the antenna can also add frequency structure on top of smooth foreground spectra. This is particularly apparent when bright foreground emissions appear in the sidelobes of the primary beam pattern \citep{Choudhuri2021, Charles2022, Trott2022, Gan2022}. Sidelobes of primary beams are typically non-smooth as a function of frequency and also not well-characterised by simulations or lab measurements, particularly in the presence of mutual coupling effects due to closely packed array configurations \citep[e.g.,][]{Fagnoni2019, Borg2020, Bolli2022a, Bolli2022b}. This inevitably leads to calibration errors, which, in turn, can also cause leakage of foreground power.

There are attempts to reduce the reliance on sky models in calibration, such as in \text{redundant calibration} \citep{Liu2010,Dillon2020}. However, redundant calibration still requires a sky model to constrain degenerate parameters that cannot be solved for using the redundant calibration equations \citep{Zhang2016,Dillon2018, Kern2020a}. Redundant calibration also assumes array redundancy, where multiple baselines that have the same orientation and length measure the same Fourier mode of sky signal. However, with actual observations, including the HERA array, baselines inevitably deviate from redundancy. This non-redundancy can be caused by multiple factors, such as antenna placement errors, antenna-to-antenna primary beam variations, and instrumental coupling \citep[e.g.][]{Orosz2019, Fagnoni2019, Kern2020a, Joseph2020, Choudhuri2021, Josaitis2022}. Non-redundancy eventually leads to foreground power leaking into the EoR window \citep{Joseph2018,Dillon2020}.  

Due to HERA's compact array layout, its observations are more sensitive to large angular structures in the sky, such as the diffuse galactic synchrotron emission \citep{Haslam1982}. Although diffuse emission models have improved over the years \citep[e.g. through the Global Sky Model - GSM);][]{Oliveira2008, Zheng2016}, 
point source catalogues remain our most well-understood calibration models, particularly with the advent of low-frequency all-sky surveys
\citep{Hurley-Walker2017, Riseley2018,Shimwell2022}.
Thus, one of the limiting factors in HERA calibration is our relatively poor understanding of the diffuse galactic foreground emission (which we will refer to interchangeably as the GSM).

In this work, we aim to improve HERA calibration by filtering out diffuse emission in the data before running calibration in order to mitigate its impact on the recovered antenna gain solutions. This idea dates back to the application of fringe rate filters for calibration in \citet{Parsons2009} and builds upon existing HERA calibration methodologies \citep{Kern2020a, Dillon2020}. We look at two specific kinds of filters: a simple high-pass filter (referred to as a ``baseline-independent notch filter'') and a baseline-dependent filter that upweights sky emission coming from within the main field-of-view of the primary beam (referred to as a "baseline-dependent main-lobe filter'').
We assess the calibration improvement by computing the reduced chi-squared after calibration and by further examining the structure of the recovered gains in Fourier space to assess the amount of residual spectral structure. Lastly, we show how our technique helps mitigate the foreground leakage in the EoR window. 

The paper is organized as follows: \autoref{sec:The_Calibration_Problem} gives an overview of the calibration problem, \autoref{sec:simulations} describes our simulations, \autoref{sec:frfiltering}
presents an overview of fringe rate filtering and an analysis of our results, and in \autoref{sec:conclusions} we present our conclusions.

%%%%%%%%%%%%%%%%%%%%%%%%%%%%%%%%%%%%%%%%%%
%%%%%%%%%%%%%%% Section One %%%%%%%%%%%%%%
%%%%%%%%%%%%%%%%%%%%%%%%%%%%%%%%%%%%%%%%%%
\section{Calibration overview}
\label{sec:The_Calibration_Problem}

\subsection{Sky-model based calibration}

When observing the sky with a radio interferometer, the electromagnetic waves from celestial sources are measured by two antennas are correlated, forming interferometric visibilities.
The measured complex-valued visibilities are related back to the intrinsic sky emission and the instrumental response of the antennas via the \emph{radio interferometer measurement equation} \citep[RIME;][]{Hamaker1996, Smirnov2011}.
We can break the RIME into two parts, one that is dependent on the position of the source emission on the sky (direction-dependent) and another that is independent of sky position (direction-independent).
The direction-dependent part of the RIME is written as
\begin{equation}
\label{eq:rime}
    V_{ij}(\nu)=\iint |A(\boldsymbol{s},\nu)|^2\, I(\boldsymbol{s},\nu) \, e^{-2\pi \imath \frac{\nu}{c} \boldsymbol{b}_{ij} \cdot \boldsymbol{s}} \, \, \frac{dldm}{n(\boldsymbol{s})},
\end{equation}
where $V_{ij}$ is the visibility formed between antenna $i$ and $j$, $A(\boldsymbol{s},\nu)$ is the average antenna primary beam response, an example of the only direction-dependent corruption that will be considered in this paper. $I(\boldsymbol{s},\nu)$ is the brightness distribution of the sky, $\nu$ is the frequency of the incoming electromagnetic wave, $\boldsymbol{b}_{ij}$ is a baseline vector connecting antenna $i$ and $j$, $\boldsymbol{s} = [l, m, n]^T$ is the position vector on the celestial sphere with an origin centred at the target field (phase centre), and $(l, m, n)$ are the direction cosines, with $n = \sqrt{1 - l^2 - m^2}$.
Any practical telescope also imparts direction-independent corruptions to the visibilities, which can be modelled using antenna-based gain terms, for example, the bandpass of the antenna front-end response. Thus, the corrupted visibilities are related to intrinsic visibilities as
\begin{equation}
    V^{c}_{ij}=g_i\, V_{ij}\, g_j^*\,  + n_{ij},
    \label{eq:corrup_model}
\end{equation}
where $V^{c}_{ij}$ are the corrupted (or measured) visbilities, $g_i$ and $g_j$ are the complex gain terms of antenna $i$ and $j$, and $n_{ij}$ is any complex thermal noise generated by the telescope.

Antenna gain calibration is the process of estimating the direction-independent gain terms ($g_i$) from the data and then correcting the data to remove their effect.
This is done by constructing a model of the true visibilities and then minimising a $\chi^2$ statistic:
\begin{equation}
    \chi^2 =\sum_{i,j} \frac{ |V_{ij}^{c}-g_i\, g_j^*V^{m}_{ij}|^2}{\sigma_{ij}^2},
\end{equation}
where $V_{ij}^{m}$ are the constructed model visibilities, and $\sigma_{ij}^2$ is the variance of the visibility, and the sum
%sum $i,j$ 
runs over all unique antenna pairs.

The chi-square minimisation depends on the accurate modelling of the intrinsic sky visibilities $V_{ij}$. It is crucial, therefore, that the sky model matches the intrinsic sky brightness as closely as possible.
This is also referred to as ``completeness'' of the sky model: if our sky model does not contain certain components of the intrinsic sky brightness distribution, then it would be ``incomplete''.
In practice, we never know the true sky distribution exactly, so all sky models are intrinsically incomplete, but the severity of this incompleteness can have different impacts on the calibration, which is the focus of this paper.

Note that self-calibration can, in principle, correct for some incompleteness effects, specifically due to phase variations in the gains and misestimated point source properties \citep{Bhatnagar2008, Eastwood2018}. However, self-calibration has shown success where we either have good priors on the nature of the underlying emission (e.g. they are point sources) or fairly accurate first guesses of the underlying sky brightness distribution. It is, therefore, not yet a reliable solution for the problems addressed in this work.

The problem of poorly modelled diffuse emission is not new, and some techniques for mitigating its impact exist.
The most common and simple approach is to exclude baselines that are most heavily contaminated by this emission in the process of calibration.
Short baselines are most sensitive to large angular scales and are, therefore, routinely thrown out prior to calibration against a sky model comprised mostly of point sources.
However, in the case of HERA, it is not possible to completely mitigate this as even longer baselines are still sensitive to diffuse emission \citep[e.g.,][]{Thyagarajan2015a, Trott2022, Mevius2022}.

Other attempts to mitigate spurious spectral structure in the calibration solutions rely on smoothing the gains across frequency \citep{Barry2016, Kern2020a} or enforcing smoothness in the fitting process \citep{Yattawatta2015, Beardsley2016}, but this does not fundamentally solve the calibration problem. Smoothing removes specific Fourier modes from the gains entirely, regardless of whether those terms held true calibration solutions or were the result of model incompleteness errors. In other words, some Fourier modes are fundamentally contaminated by model incompleteness errors, preventing us from applying calibration solutions at those scales. This is problematic if we need to invert the response of the telescope to high precision, which is generally needed for 21\,cm measurements.

\subsection{Redundant calibration}

Redundant calibration does not need a sky model to solve for a subset of the antenna gains, and obtains them by exploiting the internal redundancy of an interferometric array. Redundant baselines have the same length and orientation and, therefore, measure the same Fourier mode of the sky brightness distribution.
For example, assume we have a single baseline type, $A$, uniquely identified via its baseline vector $\boldsymbol{b}_A$. For all antenna pairs $i,j$ that share this baseline vector, the calibration equation \autoref{eq:corrup_model} from before now becomes
\begin{equation}
       V^{c}_{ij} = g_i \, V_{A} \, g_j^* + n_{ij}, 
\label{eq:red_cal}
\end{equation}
where $V_{A}$ is now a parameter of the model, called the ``redundant model visibility''.
This is repeated for all unique baseline types, eventually building up an overconstrained system of equations, which can be solved via a $\chi^2$ minimization \citep[e.g.,][]{Liu2010}:
\begin{equation}
    \chi^2 =\sum_{i,j} \frac{ |V_{ij}^{c}-g_i\, g_j^*V_{A}|^2}{\sigma_{ij}^2},
\end{equation}
where $V_{A}$ is the corresponding redundant model visibility for the $\boldsymbol{b}_{ij}$ baseline vector. Note that in this paper, the $\chi^2$ is both a function of frequency and time. The gain solutions obtained from the system of linear equations are not unique; there are at least four degenerate parameters that need to be solved for after redundant calibration: the overall gain amplitude, the model visibility amplitude and the phase gradient across the array in the east-west and north-south orientations \citep{Liu2010, Zheng2014, Dillon2018}. These parameters are, in principle, a function of polarization, frequency and time and require a sky model to be constrained.  
Thus redundant calibration itself is not entirely independent of a sky model. In this study, we will refer to this step in calibration as \text{absolute calibration} to differentiate this from \text{sky calibration} or the typical sky-based calibration discussed previously.

In the case of perfect calibration, where the sky model is complete, model visibilities only differ from the data by noise. The expectation value of chi-square $\langle \chi^2 \rangle$ is thus two times the degrees of freedom (DoF) for complex data \citep{Dillon2020}. The number of DoF, in general, is given by the difference between the number of data points and the number of fitted parameters. If we consider a single polarization and the case of sky-based calibration, the DoF are given by the difference between the number of visibilities and the number of antennas:
\begin{equation}
    {\rm DoF} = \frac{N(N-1)}{2} - N = N_{\rm bl} - N,
\end{equation}
where $N$ is the number of antennas and $N_{\rm bl}$ is the number of baselines given $N$ antennas. If we histogram the $\chi^2/{\rm DoF}$ quantity, it is expected to follow a theoretical chi-square distribution \citep{Dillon2020}. 
In the case of redundant calibration, the DoF is given by:
\begin{equation}
    {\rm DoF} = N_{\rm bl}-N_{\rm ubl}-N+2,
\end{equation}
where $N_{\rm ubl}$ is the number of unique baselines that have a single sky model visibility $V_A$, i.e. the number of redundant baseline groups \citep{Dillon2020}.

In this work, we do not simulate any non-redundancies; thus, we expect redundant calibration to work perfectly, regardless of the sky model incompleteness \citep{Dillon2018}. The required absolute calibration step that follows redundant calibration will, conversely, introduce spurious gain errors due to missing components in the sky model. 
In the remainder of the paper, we will consider $\chi^2$ statistics that include the full calibration solutions (both redundant and absolute calibration together), unlike \citet{Dillon2020}, who only plot the reduced $\chi^2$ for redundant calibration. Our calibration pipeline is a simplified version of the \texttt{hera\_cal}\footnote{\href{https://github.com/HERA-Team/hera_cal}{github.com/HERA-Team/hera\_cal}} package used for HERA observations \citep{HERA2022a}.

\section{Foreground simulations}
\label{sec:simulations}

\begin{figure}
     \centering
     \includegraphics[scale=0.6]{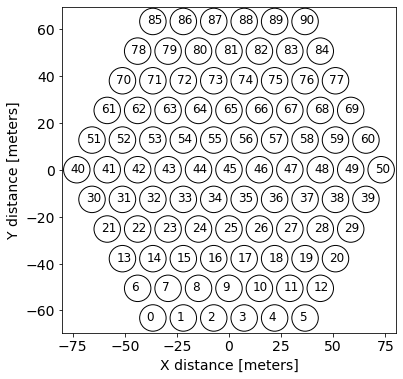}
     \caption{Simulated HERA-like array layout with 91 antennas and 14.6~meter spacing between antennas. X and Y axes indicate distances aligned along East-West and North-South directions, respectively.}
     \label{fig:array_config}
\end{figure}

\begin{figure*}
     \centering
     \includegraphics[width=1.0\textwidth]{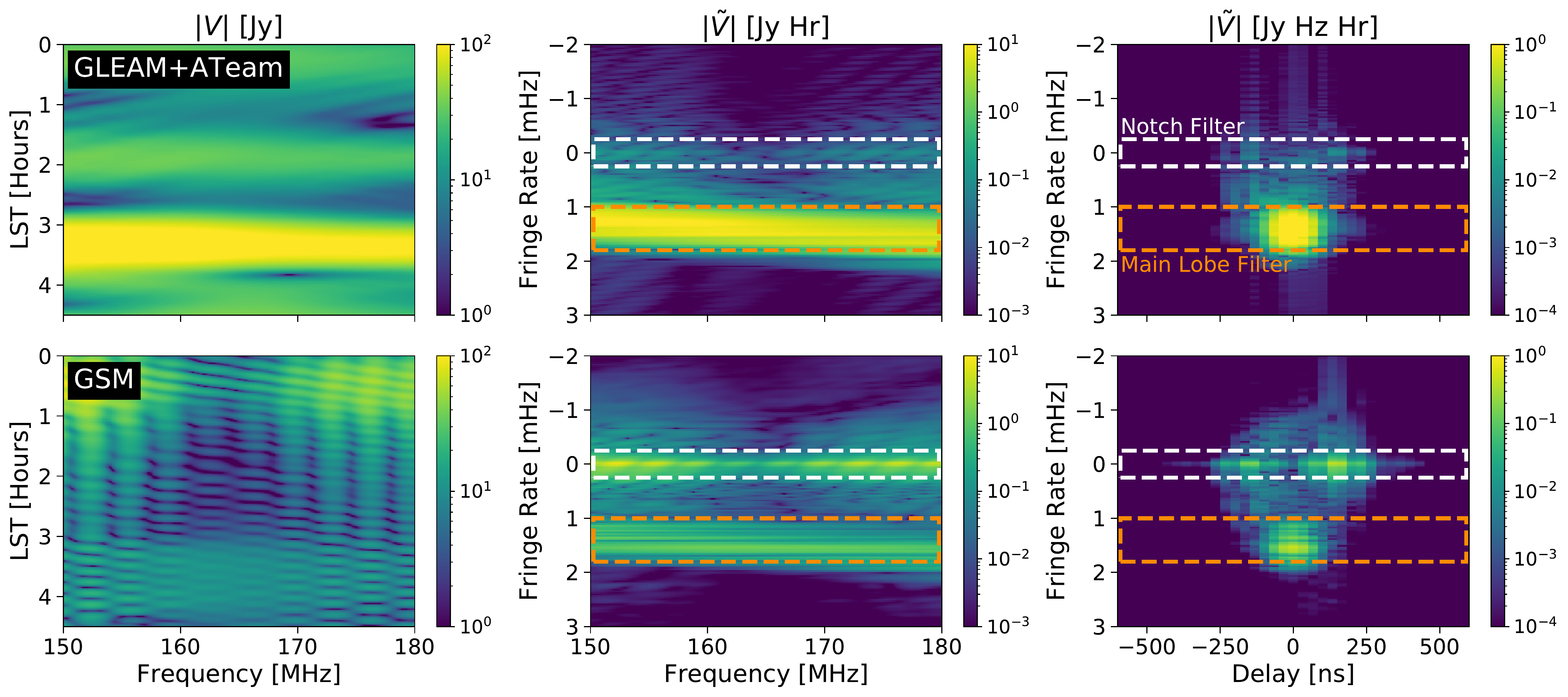}
     \caption{
     Simulated HERA visibilities of a 45~m baseline that includes the sky model ii., GLEAM sources and A-Team sources (top row) and only diffuse Galactic emission from the GSM (bottom row, see text for details). The left column shows the visibility amplitude as a function of LST and frequency, whereas the middle column amplitude of the fringe-rate visibilities as a function of fringe rate and time and the right column shows the fringe-rate, delay-transformed visibilities as a function of fringe rate and delay. 
     Point source emission is more tightly confined to low delays and non-zero fringe rates. The diffuse emission has a significant build-up of power at near-zero fringe rates due to the pitchfork effect \citep{Thyagarajan2016}.
     The boundaries of the notch (white) and main lobe (orange) filters described in \autoref{sub_sec:fringe_rate_filtering} are shown as dashed lines.
     }
     \label{fig:visibilty_plot_foregrounds}
\end{figure*}
Here we describe our simulated HERA observations.
We adopt a HERA-like array layout consisting of 91 antennas in a compact hexagonal configuration (\autoref{fig:array_config}), with a dish-to-dish spacing of 14.6 meters \citep{DeBoer2017}.
We use an electromagnetic simulation of the farfield HERA Phase I dish and dipole feed response (not including the effects of cross-coupling) to model the antenna primary beam \citep{Fagnoni2019}.
In order to simulate a quasi-realistic case, we assume the following representation of the gain $g$ for the $j$th antenna:

\begin{equation}
    g_j(\nu)=A_j (\nu) e^{i \phi_j(\nu)},
\end{equation}
 where the amplitude $A$ follows a frequency power law 
\begin{equation}
    A_j(\nu)=A_j\,  \left( \frac{\nu}{ {\rm150~MHz}} \right)^{b_j},
\end{equation}
and $A_j$ and $b_j$ are drawn for each antenna from a Gaussian distribution $\mathcal{N} \sim (\bar{x}_{A}=0.30, \sigma_{A}=0.001)$ and $\mathcal{N} \sim (\bar{x}_b=-2.6,\sigma_b=0.2)$, respectively. The phase $\phi_i(\nu)$ is modeled as:
\begin{equation}
  \phi_j(\nu)=\sin{(w_a \, \nu)} + \cos{(w_b \, \nu)},
\end{equation}
where $w_a$ and $w_b$ are drawn for each antenna from a Gaussian distribution $ \mathcal{N} \sim (\bar{x}_w=0.005, \sigma_w=0.0005)$. The mean values of the parameters $A$ and $b$, as well as their standard deviation $\sigma_A$ and $\sigma_b$, are informed by the gain solutions from actual HERA observations \citep{Kern2020a}. The mean phase variation $\bar{w}$ is based on single antenna phase dependency of actual HERA gains where the cable delay and geometric phase offset have been removed, and the variation of the mean phase between antenna stations, i.e. $\sigma_w$, is chosen to be within $10\%$. This exact choice does not strongly affect the results discussed below, but is a reasonable starting point for performing calibration after an initial delay calibration step.

We simulate HERA visibilities for three distinct sky models: i. shallow GLEAM + A-Team model; ii. deep GLEAM + A-Team model, and iii. GSM diffuse model. The GLEAM catalogue is a point source catalogue covering the HERA field of view considered in this work \citep{Hurley-Walker2017}. The shallow GLEAM catalogue includes GLEAM sources only down to a flux density limit of 100~mJy (at 180~MHz), while the deep GLEAM catalogue goes down to a flux density limit of 25~mJy. In both cases, we also add models for bright, extended radio galaxies that we label as the A-Team \citep[Fornax~A, Hydra~A, Pictor~A, Hercules~A, Virgo~A, Cygnus~A, Cassiopeia~A;][]{MCKinley2015, Byrne2022}. Lastly, we also include the diffuse emission from the GSM \citep{Zheng2016}. 

For each of the three sky models,  we use \texttt{healvis} \citep{Lanman2019b} to simulate HERA visibilities using the array layout in \autoref{fig:array_config} and numerically evaluate \autoref{eq:rime}, assuming the entire observable hemisphere as the field of view. For the primary beam response, we use an electromagnetic simulation of the Phase I HERA dipole antenna response \citep{Fagnoni2019}.
Our simulations span $150 - 180$~MHz in frequency with a 97~kHz channel resolution, and are made for local sidereal times between $-2^{\rm h} < {\rm LST} < 7^{\rm h}$ range at a 60~s cadence, corresponding to the time when the Galactic center falls partially in the far sidelobes.

We form two separate sets of visibilities, i.e. a mock calibration model that includes only the shallow GLEAM and A-Team components and a ``true'' mock dataset that includes all three sky models. Finally, we apply the simulated gains to the visibilities corresponding to the true dataset and added Gaussian thermal noise at a level comparable to HERA Phase I measurements, with a signal-to-noise ratio of $\sim 100$ per visibility \citep{HERA2022a}.

\autoref{fig:visibilty_plot_foregrounds} shows an example of the deep GLEAM + ATeam (top row) and the GSM (bottom row) visibility data product for a 45~meter baseline. It highlights the different spectral and temporal structure of the two sky components in the native telescope measurement space (frequency vs time, left panels), as well as in a 1D and 2D Fourier space representation that we discuss in \autoref{sec:frfiltering}. The dashed lines show the extent of the two kinds of filters that we apply to the data, defined in detail in \autoref{sec:frfiltering}.
This clearly illustrates how the two models manifest in the different spaces and hints at how we may be able to isolate them, at least partially, by applying suitable filters.

\section{Fringe Rate Filters for Improving Calibration}
\label{sec:frfiltering}

Here we demonstrate how time-based filtering (what we call fringe-rate filtering) can improve the data-to-model alignment and thus partially mitigate some of the erroneous spectral features caused specifically by poorly modelled diffuse emission\footnote{This technique can also be extended to suppress similar spurious features in the recovered gains created by non-redundancy in the data (e.g. from crosstalk or per-antenna beam errors), which we defer to future work.}.
First, we discuss the Fourier representations of the data and then discuss calibrating our mock HERA data with and without the two different types of fringe rate filters.

\subsection{Fourier representations}
\label{sec:fourier_space}

\begin{figure*}
    \centering
    \begin{tabular}{cc}
         \includegraphics[scale=0.6]{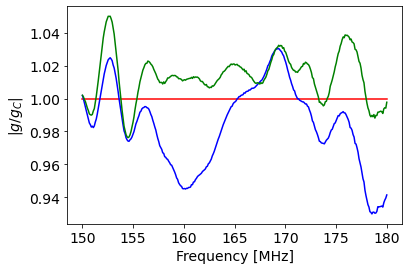}
         \includegraphics[scale=0.6]{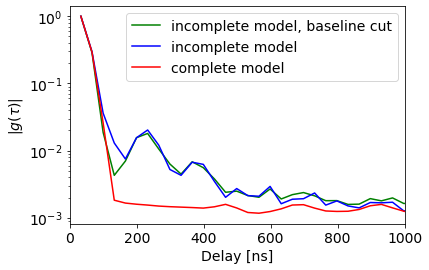}  
    \end{tabular}
    \caption{Recovered gain solutions after calibrating with an incomplete sky model. Left: The amplitude ratio $g/g_c$ between the gain $g$ obtained from the three simulated cases and the gain $g_c$ obtained from calibration when a complete sky model is assumed. Both terms are the average gain across all the antennas at  LST~$=1.2^{\rm h}$. Red is the case of a complete sky model (which results in a ratio of one), blue is for the incomplete sky model, and green is for the incomplete sky model with a 40~m minimum baseline cut. Right: Here we show the amplitude of the gains in delay space after averaging over all antennas. The excess gain structure as a function of frequency due to the sky model incompleteness is evident even when a baseline cut is adopted (the bumps for $\tau>200$ ns). This excess ``shoulder'' is particularly visible in delay space and its profile is fairly similar in both cases where the sky model is incomplete (with or without a baseline cut).
    For delays $\tau>200$ ns this spurious structure is at its peak over an order of magnitude above the noise floor from the complete-model-derived gains (red).
    }
    \label{fig:gains_incomplete}
\end{figure*}

The native measurement space of the interferometric visibilities is observing frequency ($\nu$) and local sidereal time (LST).
For telescopes that operate in drift-scan mode, the local sidereal time is effectively just the right ascension directly overhead at any given time. We refer to the Fourier dual of frequency as the delay domain, formed by taking a Fourier transform of the visibilities across the frequency axis:
\begin{equation}
\label{eq:delay_transform}
    \tilde{V}(\tau) = \int V(\nu) \, \e^{-2 \pi i \tau \nu} d\nu,
\end{equation}
where $\tau$ is a delay (in seconds) and $\tilde{V}(\tau)$ is the Fourier pair of the frequency; similarly, we can also define delay-transformed antenna gains to be the Fourier transform of the antenna gains along the frequency axis:
\begin{equation}
\label{eq:delay_transform_gain}
    \tilde{g}(\tau) = \int g(\nu) \, \e^{-2 \pi i \tau \nu} d\nu.
\end{equation}
We refer to the Fourier dual of time (or LST) as the \emph{fringe rate} and the fringe-rate visibility as the Fourier transform of the visibility along the time axis:
\begin{equation}
\label{eq:fr_transform}
    \tilde{V}(f) = \int V(t) \, \e^{-2 \pi i f t} dt,
\end{equation}
where $f$ is the fringe rate (in units of Hz)\footnote{Referred to as delay-rate by \citet{Parsons2009}, but subsequently called fringe rate in works thereafter.}.

The Fourier space of the visibilities (delay and fringe rate) is useful for separating different signals in the data. Delay space separates signals based on their incident angle from the pointing of the telescope, and is the basis for the delay spectrum foreground avoidance technique \citep{Parsons2012a, Liu2014a, HERA2022a}.
Similarly, the fringe rate basis (for drift-scan observations) also separates signals based on their relative motion through the fixed interferometric fringes, acting as another form of separation of signals on the sky \citep{Parsons2009, Parsons2016}.

 A fringe rate filter can, therefore, be used to suppress sky signals that appear at specific fringe rate values, for instance, outside the primary beam field of view (\autoref{fig:visibilty_plot_foregrounds}). \citet{Parsons2016} indeed describe how sky signals within the field of view appear at specific, baseline-dependent fringe rates. In contrast, sources outside the field of view appear to have a fringe rate that is closer to zero (i.e., a near-constant time response). In other words, fringe rate filtering is equivalent to ``sculpting'' the primary beam. We refer the reader to  \citet{Parsons2016} for a more detailed description of fringe rate space for drift-scan observations.
 
Fourier representations have been used widely to separate signals and systematics in 21\,cm observations \citep[e.g.][]{Kolopanis2019, Kern2020b, Josaitis2022}; however, these have generally been applied to pre-calibrated data. Here we explore using these basis to separate signals before calibration.

\subsection{Calibration Without Filtering}
\label{sec:calibration_wo_filtering}

To gain an intuition for how calibration would work without
applying any visibility filtering, we ran three different kinds of simulated observation cases. First, we used a complete sky model, i.e. the same components used for the simulations, the deep GLEAM catalogue, the A-team sources and the GSM to calibrate raw data. In the second case, we calibrated by using an incomplete sky model, i.e. including only the shallow GLEAM catalogue and the A-team sources. In the third case, we use the same sky model as the second case, but with the addition of a baseline cut at 40~m, i.e. all the baselines shorter than 40~m were excluded before calibration. This is a commonly used approach in order to mitigate the contribution of diffuse emission that appears mostly on short baselines.

The recovered gains in each case are shown in \autoref{fig:gains_incomplete}.
The gains recovered in the complete sky model case represent the true simulated gains, as they only differed from the true gains by noise. Notice this is not the case for the incomplete sky model case. There is an excess frequency structure that is caused by the unmodelled foreground emission, coupled with the sidelobe primary beam response. During calibration, gains absorb a portion of this unmodelled term, resulting in increased frequency structure. This indeed has been shown in literature, in both simulations and observations \citep[e.g.,][]{Wijnholds2016, Barry2016, Ewall-Wice2017, Byrne2019, Kern2020a}. Notice that there is no appreciable difference in the gains whether a baseline cut is adopted or not. The frequency structure caused by an incomplete sky model introduces excess power in the gains at delays in the range $100 < \tau < 400$~ns, right where the sensitivity to the EoR signal is the highest.

\begin{figure}
    \centering
    \includegraphics[width=1.0\linewidth]{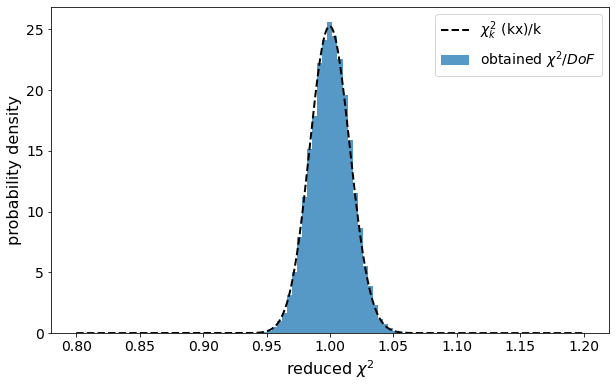}
   \caption{Histogram of the reduced $\chi^2$ (blue), i.e. $\chi^2/{\rm DoF}$, obtained when a complete sky model is used for calibration. The dashed line is the theoretical reduced $\chi^2$  distribution with $k=2 \times {\rm DoF}$, i.e. the chi-square distribution degrees of freedom, and $x = {\chi^2/\rm DoF}$. This is a simple demonstration that our calibration pipeline works as expected in the limit of a complete sky model. In other words, our gains are accurate down to the expected thermal noise.}
    \label{fig:chisquare_redundant_cal}
\end{figure}

To demonstrate that our calibration pipeline works as intended
in an ideal case, we plot the reduced $\chi^2$ obtained from the visibilities calibrated with the recovered gains in the complete sky model case (\autoref{fig:chisquare_redundant_cal}). The reduced $\chi^2$ of the calibrated data shows good agreement with a theoretical unit-$\chi^2$ distribution (dashed line), showing that our calibration pipeline is both unbiased and recovers the gains down to the thermal noise floor in the ideal circumstance of a complete sky model. 

\begin{figure*}
    \centering
      \begin{tabular}{cc}
      \includegraphics[scale=0.55]{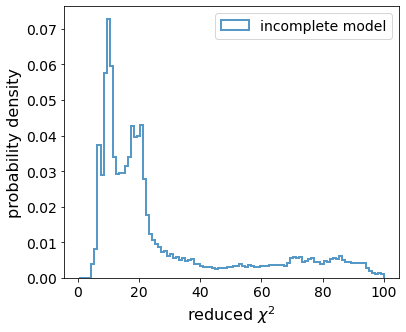}
      &
          \includegraphics[scale=0.55]{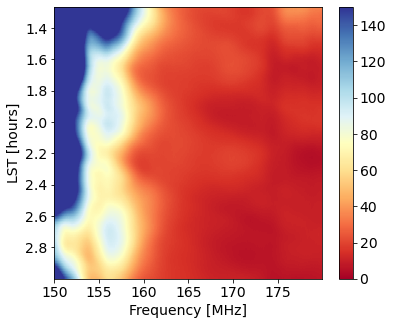}   
      \end{tabular}
    \caption{The resultant reduced chi-square after calibrating with an incomplete sky model and including a 40-meter minimum baseline cut. Left: Histogram of the reduced chi-square with baseline (blue). Right: The reduced chi-square as a function of time and frequency. The missing components in our sky model leads to a clear bias in the reduced chi-square after calibration. The increase at low frequencies is attributed to the increased signal-to-noise ratio in the simulated visibilities.}
    \label{fig:chisqaure_incomplete}
\end{figure*}

The bias introduced by the sky model incompleteness, with a baseline cut is evident in the reduced $\chi^2$ distribution (\autoref{fig:chisqaure_incomplete}), which shows a strong deviation from the theoretical one. The mean of the reduced $\chi^2$ distribution is biased toward high values, with a mean of $\chi^2/{\rm DoF} \sim 19$, and a tail extending up to $\chi^2/{\rm DoF} \sim 100$. \autoref{fig:chisqaure_incomplete} also shows the $\chi^2$ distribution as a function of time and frequency spanning LSTs of $1.3 - 3.0$~h (right panel). The reduced $\chi^2$ varies noticeably as a function of frequency and LST; in particular, its bias is higher at low frequencies where the diffuse emission is significantly brighter.
The reduced $\chi^2$ is less biased around LST~$\sim 2.8^{\rm h}$, where bright, compact sources dominate the sky emission yielding to a more complete sky model.

\subsection{Fringe Rate Filtering Before Calibration}
\label{sub_sec:fringe_rate_filtering}

Here we propose to fringe rate filter the data before calibration as a way to improve the data-to-model match and thus mitigate some of the effects seen in \autoref{sec:calibration_wo_filtering}.
Note that fringe rate filtering is applied to \emph{both} the raw data and the sky model visibilities in the same manner for each of the different filters described below.
Also, note that the derived gains are applied to the unfiltered visibilities.
In other words, our fringe rate filtering is only used en route to derive gains that are more robust to an incomplete sky model and are not used downstream in our analysis (thus mitigating concerns of a possible cosmological signal loss).

Note that, for the two filters described below, we apply them to the data using the DAYENU filtering formalism described in \citet{Ewall-Wice2020}, which relies on the Discrete Prolate Spheroidal Sequences \citep[DPSS;][]{Slepian1978}.

\subsubsection{Notch Filters}
\label{sec:notch_filters}
\begin{figure*}
    \centering
    \includegraphics[width=1\textwidth]{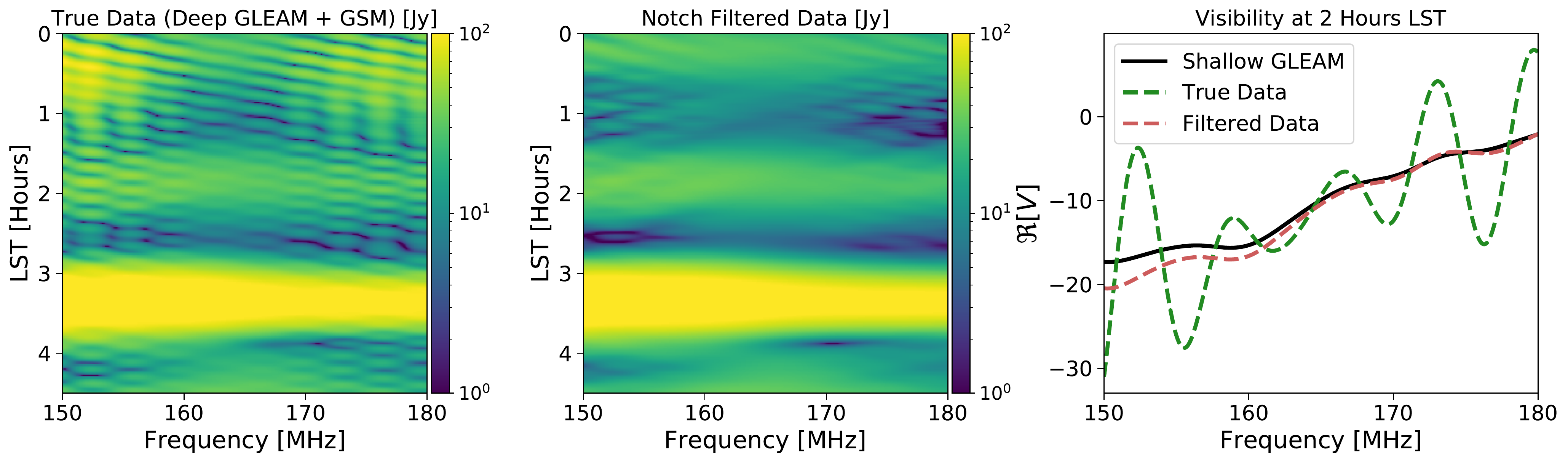}
    \caption{Simulated visibility amplitude as a function of frequency and LST for a 45-meter HERA baseline. The left panel includes the deep GLEAM model and the GSM. Same is shown at the centre, but after applying a 0.25~mHz notch fringe-rate filter. The right panel shows the real part of the visibility at an LST~$ = 2^{\rm h}$, for the unfiltered (dashed green line) and filtered case (dashed red line), alongside the shallow GLEAM model (black) used for calibration. After fringe-rate filtering, visibilities are in much better agreement with the model, with the large frequency ripples significantly suppressed.
    }
    \label{fig:fr_filters}
\end{figure*}

We first consider a symmetric baseline-independent notch filter $F(f)$ centred at $f = 0$~mHz fringe rate,  i.e. a high-pass filter, defined as 
\begin{equation}
F(f) = 
\left\{
    \begin{array}{lr}
        10^{-8}, & |f| \leq f_{\rm max}\\
              1, & |f| > f_{\rm max}
    \end{array}
\right\},
\end{equation}
with 
%three different half-widths of 
$f_{\rm max} = [0.25,0.40,0.60]$~mHz, respectively. We refer to the three filters as $f_{25}$, $f_{40}$ and $f_{60}$, respectively. 
The bounds $f_{\rm max}$ of the $f_{25}$ filter are the white dashed lines shown in \autoref{fig:visibilty_plot_foregrounds} and indicate the breadth of the emission suppressed in fringe rate space.
\autoref{fig:fr_filters} shows the impact of baseline-independent  notch filters on foreground emission in time and frequency space. As a result of filtering out some of the diffuse foreground component, the visibility spectra of
the filtered data more closely match that of compact source sky model.

\subsubsection{Main Lobe Filters}
\label{sec:main_lobe_filters}

We also considered a second type of filter, which we refer to as a ``main lobe'' filter, because it aims to suppress emission from outside the primary beam field-of-view.
In contrast to the baseline-independent notch filter, which only filters out emission in a region near $f\sim0$~mHz, the baseline-dependent main-lobe filter suppresses the signal everywhere \emph{except} near the peak emission of the sky model in fringe rate space (\autoref{fig:visibilty_plot_foregrounds}).

In principle, the baseline-dependent main-lobe filter should be a frequency-dependent filter, but because we operate over a relatively small bandwidth (30~MHz), we approximate it as frequency-independent, with little impact on our final results.
Note that the response of the filter within the pre-defined fringe rate bounds is uniform, such that it can be thought of as a top-hat filter (similar to the baseline-independent notch filter).
The filter presented here is similar to the filter described in \citet{Josaitis2022} for understanding mutual coupling. 

\begin{figure}
    \centering
    \includegraphics[width=.45\textwidth]{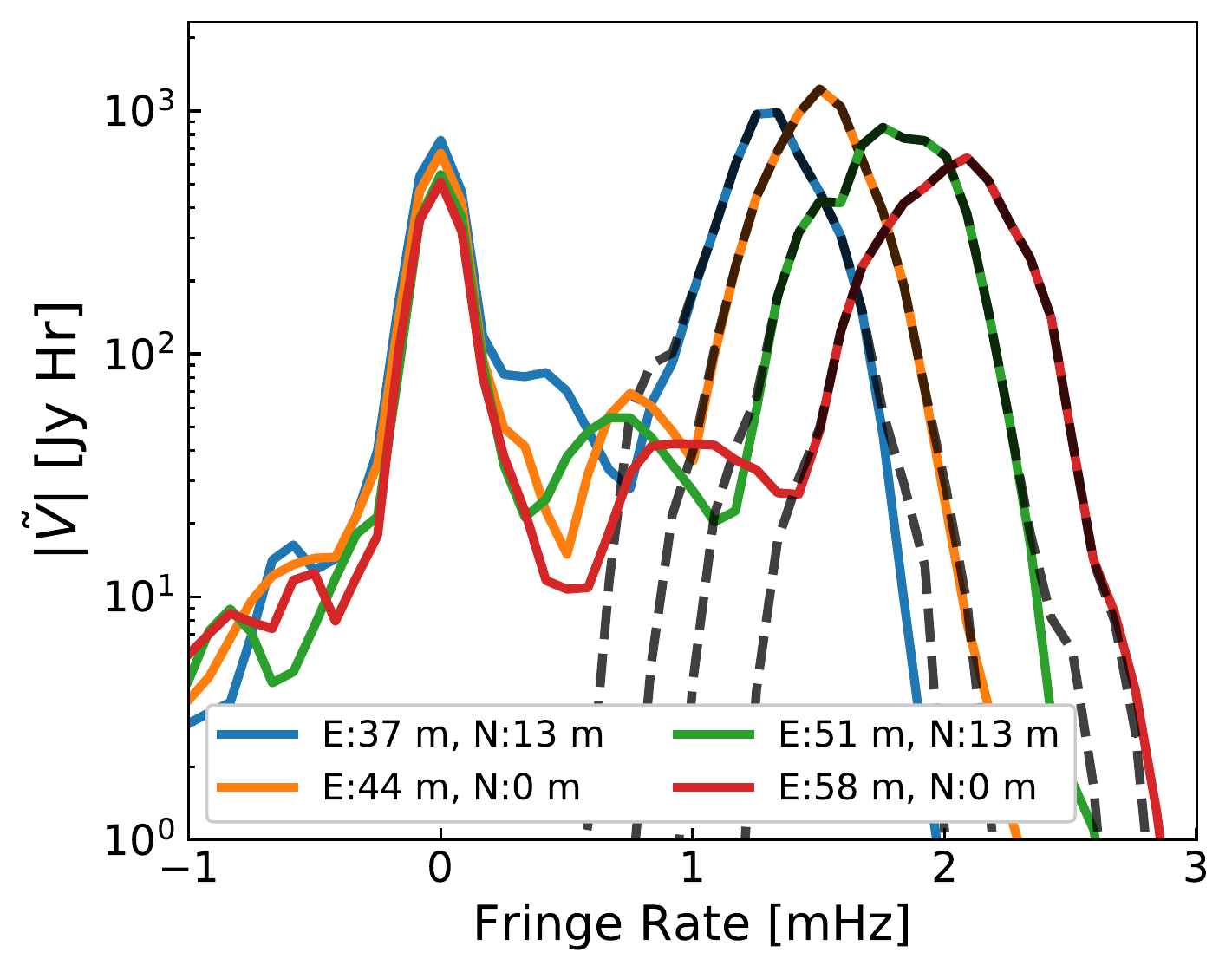}
    \caption{Model visibility amplitude (GSM~+~GLEAM~+~A~Team) as a function of fringe rate for a few different baselines in the HERA array with a projected East-West component greater than 30~meters (\autoref{fig:array_config}). Solid (dashed) lines indicate visibilities before (after) applying the baseline-dependent main-lobe filter, showing that the filter retains only the emission at positive fringe-rate values associated with the primary beam. The legend marks the East and North extent of each baseline vector in meters.}
    \label{fig:mainlobe_filter_applied}
\end{figure}

\begin{figure}
\centering
\includegraphics[width=\linewidth]{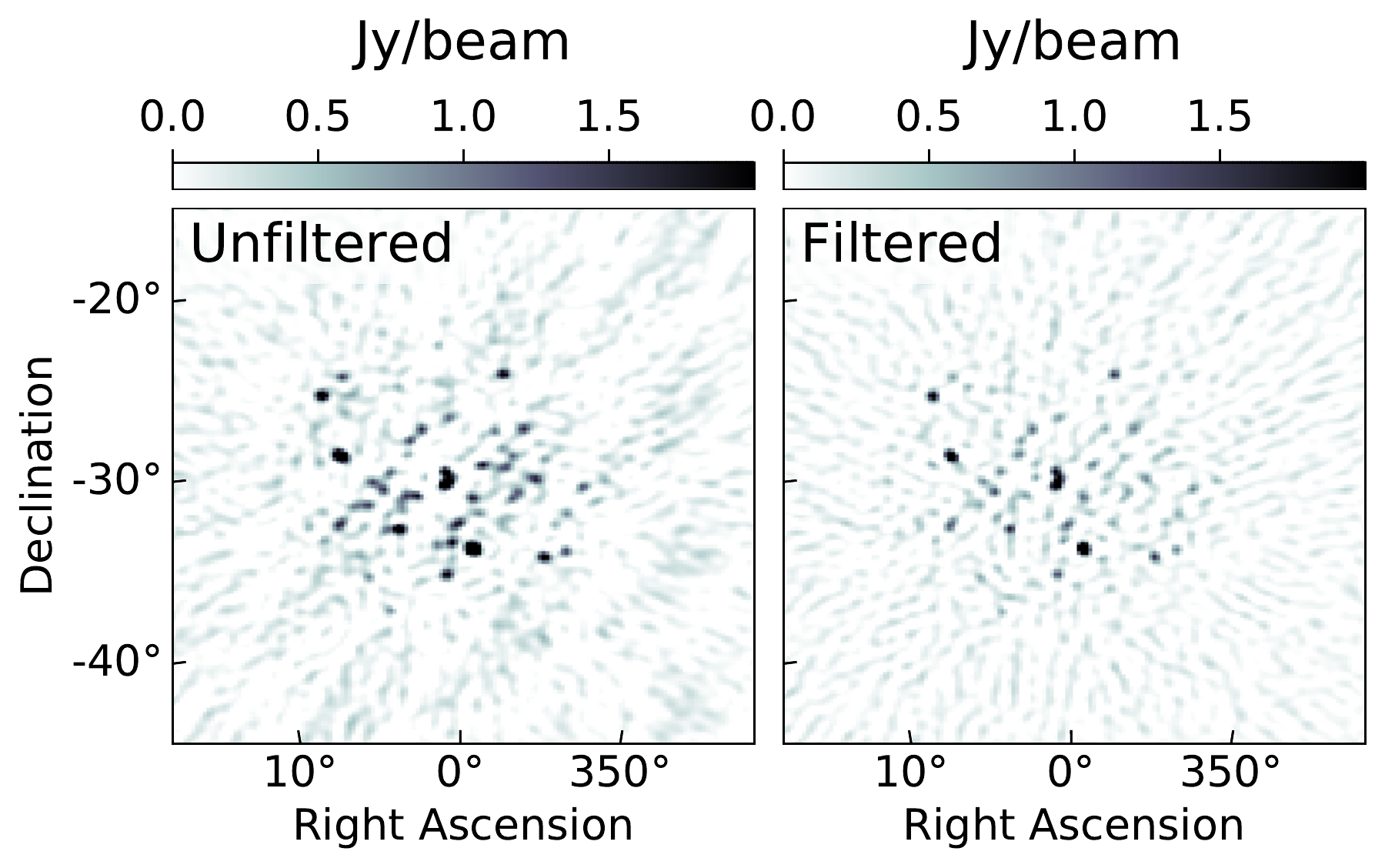}
\caption{Simulated dirty image that includes the GLEAM catalogue, the A-team sources and the GSM before (left) and after (right) applying the baseline-dependent main-lobe filter. The simulated observation is 10~minute long and centred at $\alpha=0^\circ$. The apparent absence of sky emission outside the central $\sim 10^\circ$ area is due to the primary beam attenuation. The patchy structure in the left panel is the sidelobe structure from Galactic emission located outside the primary beam, which is largely removed after filtering.}
\label{fig:widefield_maps}
\end{figure}

The bounds of the baseline-dependent main-lobe filter are determined by its centre $f_0$  and its half-width $f_w$. These parameters are determined, for each baseline, by fitting a Gaussian profile $G(f)$ to the sky model visibilities in fringe rate space:
\begin{equation}
    G(f) = A \, e^{-\frac{(f - f_{0})^2}{2 \sigma^2}},
\end{equation}
where $A$ is the amplitude of the Gaussian, $\sigma$ is its standard deviation, and $f_0$ is its mean.
After the fit, we set the main lobe filter centre to be $f_0$ and its half-width to be $f_w = 2\sigma$, such that the full width of the main lobe filter is $4\times$ the fitted Gaussian's standard deviation. \autoref{fig:mainlobe_filter_applied} shows the foreground visibilities for a number of selected baselines after applying the main lobe filter. The filter not only removes emission near $f\sim0$ mHz, but also removes all emission that is not centrally peaked at positive fringe rates.
The impact of the filter can also be seen clearly in image space (\autoref{fig:widefield_maps}), where patchy structure coming from Galactic emission in the primary beam sidelobe is largely suppressed while compact source emission is retained.


\subsubsection{Computing the Power Spectrum}
\label{sec:power_spectrum}

In addition to the reduced $\chi^2$ and the delay transform of the gains, we also used the power spectrum as a metric to assess the impact of fringe rate filters on calibration. We compute the per-baseline power spectrum of the visibilities, $P(\tau, b)$, following the delay approximation \citep{Parsons2012a}.
This simple formalism relates the Fourier transformed visibilities of each baseline, $\tilde{V}_b(\tau)$, directly to the 21\,cm power spectrum
\begin{equation}
    P(\tau, b)=|\Tilde{V}_b(\tau)|^2 \, \bigg(\frac{\lambda^2}{2k_B}\bigg)^2\bigg(\frac{D_c^2\Delta D_c}{B_{\rm eff}}\bigg)\bigg(\frac{1}{\Omega B_{\rm eff}}\bigg),
    \label{eq:Power_spectrum_visibilities}
\end{equation}
where $\lambda$ is the center wavelength of observing bandwidth, $k_B$ is the Boltzman constant, $B_{\rm eff}$ is the effective bandwidth, $D_c$ is the comoving distance at the redshift of our measurement, $\Delta D_c$ is comoving distance parallel to line of sight, $\Omega$ is the field of view solid angle, $\tilde{V}_b$ is the Fourier transformed visibility, and $b$ is the visibility baseline length.
We can map $\tau$ to the line-of-sight cosmological Fourier wavevector $k_{\parallel}$ using the relation \citep{Thyagarajan2013}:
\begin{equation}
    k_{||} = \frac{2\pi \nu_{21} H_0 E(z)}{c(1+z)^2} \, \tau,
\end{equation}
where $\nu_{21} = 1420$ MHz, $H_0$ is the Hubble constant, $E(z)=[\Omega_m(1+z)^3+\Omega_\Lambda]^{1/2}$, $\Omega_m$ is the normalized matter content and $\Omega_\Lambda$ is the normalized dark energy content.
Note that while the baseline length maps to the perpendicular Fourier wavevector ($k_\perp$) and thus contributes to the overall $|\bf k|$ magnitude, for short baselines its contribution is negligible and we will drop it for simplicity and simply quote $P(k_\parallel)$ hereafter.

\subsubsection{Summary}

The two filters described above both aim to limit the influence of the poorly modelled diffuse emission (GSM) component in the data. We can readily see that a significant amount of power from the GSM is found at and around $f\sim0$ mHz, although there is still a non-zero amount of power at positive fringe rates.
The baseline-independent notch filter acts to remove emission at zero and near-zero fringe rates, thus improving the prominence of the point sources in the visibilities.
The baseline-dependent main-lobe filter, on the other hand, filters out everything 
outside the main lobe, i.e. a larger amount of signal and noise.

To gain more intuition as to why these fringe rate filters not only improve the data-to-model match but also \emph{reduce spurious spectral structure in the recovered gains} we plot the pre-filtered and post-filtered true visibility in \autoref{fig:fr_filters}, and compare them to the incomplete model visibility used for calibration.
We see that the unfiltered true dataset has large ripples (green dashed) that is not reflected in the sky model (black solid); however, after applying a fringe rate filter (red dashed), the true data is brought into closer alignment with the sky model. Specifically, the large ripples that are in the true data but absent in the sky model are greatly suppressed.
We will see in the next section that this improvement in the data-to-model match translates to smaller errors in the gain solutions at the Fourier scales of the ripple.

\subsection{Effects of fringe rate filters on calibration}
\label{sec:fr_effects}

Before we consider the application of fringe rate filters, we first discuss the side effects that fringe rate filters may have on visibilities. After filtering, the visibility noise is suppressed and also becomes correlated from time-to-time. Both of these will affect the computed reduced $\chi^2$: the noise correlation can be corrected by adjusting the degrees of freedom (DoF), whilst the signal loss by adjusting the weights $\sigma_{ij}$. In this study we will only consider adjusting the weights to correct for the reduced noise amplitude, as adjustments to the DoF do not strongly impact the results of the study.

White thermal noise is uncorrelated in the visibilities and thus occupies all Fourier modes in the data uniformly. 
The noise amplitude will in theory depend on the frequency and observing time, but for our studies, we assume that the noise is both time and frequency independent.
Thus we can use a single number to describe the noise in the visibilities, which is its total variance at each time and frequency pixel, $\sigma^2$.
A fringe rate filter suppresses noise depending on how many Fourier modes it removes from the data \citep{Ali2015, Parsons2016}.
The resulting noise variance after filtering is related to the integral of the filter shape, $F(\cdot)$, in fringe rate space:
\begin{align}
\sigma_f^2 = \sigma^2\frac{\int F(f)\phi(f) df}{\int \phi(f)df},
\end{align}
where $\phi(f)$ is a binary windowing function that isolates the fringe rates measured in the data (i.e. 1 if $f_{\rm min}<f<f_{\rm max}$ otherwise 0), and $\sigma_f^2$ is the pixel variance after filtering.
When computing the $\chi^2$ after filtering, we use this updated noise variance.

When we filter the raw data with a chosen fringe rate filter, we also apply the same filter to the model. This way, we ensure that the impacts of the filter are consistent between the data and the model. The chi-square for the filtered data is then given by:
\begin{equation}
    \chi^2 =\sum_{i,j} \frac{ |V_{ij}^{d,f}-g_i\, g_j^*V^{m,f}_{ij}|^2}{\sigma_{f,ij}^2},
\label{eq:chi_square_filtered}
\end{equation} 
where $V_{ij}^{d,f}$ and $V_{ij}^{m,f}$ are filtered data and model visibilities respectively.

\subsection{Calibration with the notch filter}
\label{sec:applying_notch_filters}
\begin{figure}
    \centering
         \includegraphics[scale=0.55]{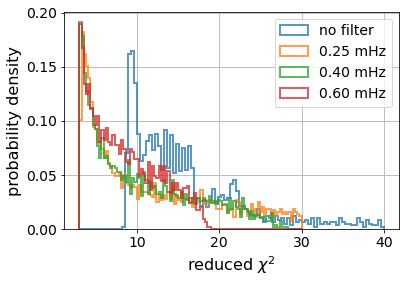}
    \caption{Histogram of the reduced chi-square obtained from calibration with an incomplete sky model with and without our fringe rate baseline-independent notch filters for field centred at LST $=2^h$. Colours indicate the different filters used, i.e. $f_{25}$ (orange), $f_{40}$ (green), $f_{60}$ (red) and no filter (blue). The reduced $\chi^2$ improves significantly after the use of a baseline-independent  notch filter, both in terms of its peak value being closer to one, and a shallower tail at high values.}
\label{fig:chisqaure_complete_vs_incomplete_filtered}
\end{figure}
\begin{figure}
    \centering\includegraphics[scale=0.55]{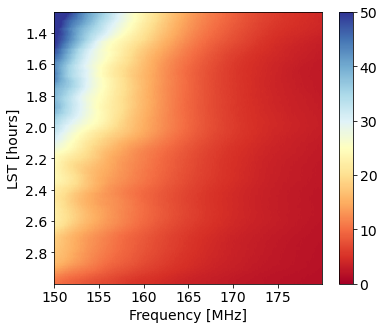}
    \caption{Reduced chi-square as a function of frequency and LST computed from calibrated visibilities after filter $f_{40}$ is applied. A much better agreement between the data and model (i.e. lower reduced $\chi^2$) is achieved compared to the case of no filtering (\autoref{fig:chisqaure_incomplete}).}
    \label{fig:chisqaure_with_filter}
\end{figure}
\begin{figure*}
    \centering
    \includegraphics[scale=0.55]{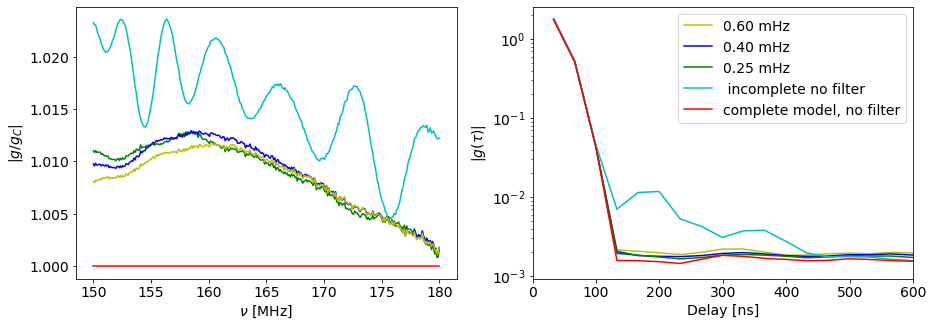}
    \caption{The recovered gains after calibration (similar to \autoref{fig:gains_incomplete} but for a field centred at LST=$2^h$). Here we plot the gains derived from a complete model (red), an incomplete model with no filtering (cyan), and an incomplete model having applied baseline-independent  notch filters (\autoref{sec:notch_filters}) of increasing width (green, blue, yellow). The significant amount of spectral structure seen when calibrating against an incomplete model (black) is heavily suppressed after applying the baseline-independent notch filters. All filters explored seem to have effecively the same performance.}
    \label{fig:gains_filtered_data}
\end{figure*}
\begin{figure*}
    \centering
    \begin{tabular}{c} 
       \includegraphics[scale=0.5]{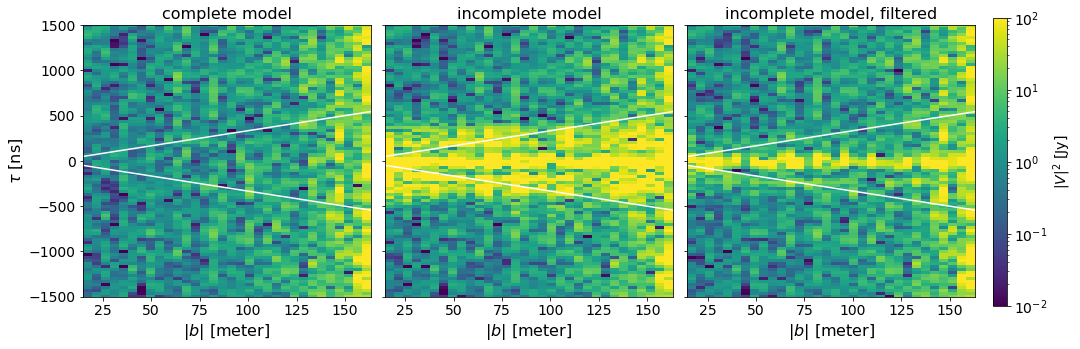} 
    \end{tabular}
    \caption{The residual of the squared visibilities in delay space as a function of baseline length. We average all baselines of equal length (similar to the commonly used power spectrum ``wedge'' plots in $k_\parallel-k_\perp$ space).
    The residual is taken with respect to the noise-free complete visibility model.
    We show residuals in the case where calibration assumes a complete sky model and is thus perfect, leaving only noise in the residual (left panel), an incomplete sky model (middle panel) and a $f_{25}$ notch filtered scenario (right panel). The visibilities are taken from a single time integration at LST~$=1.2^{\rm h}$. The white line marks the horizon limit of the baselines, which bounds the natural extent of foreground emission in the data. Note that the increased noise amplitude with increasing baseline length comes simply because there are fewer longer baselines than short baselines in the data. The right panel shows significant suppression of foreground leakage due to gain errors, as expected.}
    \label{fig:power_spectrum_wedge}
\end{figure*}
\begin{figure*}
    \centering
    \begin{tabular}{cc}
      \includegraphics[scale=0.55]{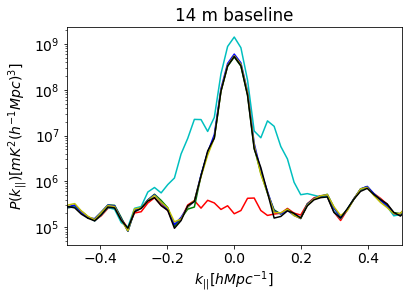}   & \includegraphics[scale=0.55]{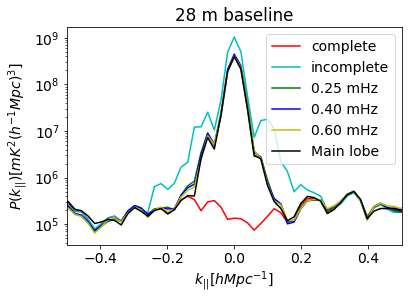}

    \end{tabular} 
    \caption{The residual delay power spectrum of the calibrated raw visibilities with respect to the complete model (noise-free) visibilities, averaged over redundant baselines and 8 time integrations centred at LST~$=2^{\rm h}$. Left: A 14~m redundant baseline group, calibrated using a complete (red) and incomplete (cyan) sky model, and having applied a the $f_{25}$ notch filter (green), the $f_{40}$ notch filter (blue), $f_{60}$ notch filter (yellow) and the baseline-dependent main-lobe filter (black) prior to calibration. Right: Same, but for the 28~m redundant baseline group. Note that for these short baselines, the $k_\perp$ contribution is negligible to the total $\boldsymbol{k}$ magnitude, such that $k_\parallel$ is effectively equivalent to the total magnitude $|\boldsymbol{k}|$.}
    \label{fig:power_spectrum_all_filter}
\end{figure*}

We first investigate calibration after applying the baseline-independent  notch filter, which suppresses emission centred at zero fringe rates. For baselines with small East-West projected length, however, emission from the main lobe appears near to zero fringe rate too \citep{Parsons2016}. For instance, the main lobe peak occurs at a fringe rate $< 0.8$~mHz for a baseline with an East-West projection length smaller than 14~m. After filtering, the SNR for short baselines is reduced as the filter also removes relevant foreground emission.
We, therefore, excluded from calibration all baselines with E-W projection shorter than 30~m to circumvent this issue, resulting in a 46\% baseline loss for our simulated HERA configuration. We will later see that the benefits of applying the filter outweigh the sensitivity loss.

Not all the LST intervals are ideal for calibration; typically, fields where compact sources are prominent, are better suited for calibration. Here we focus on the LST~$=2^{\rm h}$ interval, the typical LST range used for HERA calibration. In this LST range, bright, compact sources dominate the sky emission, and the bias due to an incomplete sky model is less pronounced. We considered a total of eight integration times around LST~$=2^h$. \autoref{fig:chisqaure_complete_vs_incomplete_filtered} shows the chi-square obtained from the calibration with an incomplete sky model and after applying the baseline-independent notch filter (equation~\ref{eq:chi_square_filtered}; the same case as \autoref{fig:chisqaure_incomplete}). Compared with the case of unfiltered visibilities, the $\chi^2$ is now less biased, with a mean value of 12, 9 and 6.5
%12.1, 9.1 and 6.5 
for filters $f_{25}$, $f_{40}$ and $f_{60}$, respectively, compared to 48
%48.1 
for the unfiltered visibilities. The wider the filter, the larger the suppressed emission, leading to a better match between the intrinsic and sky model visibilities. 

As mentioned in \autoref{sub_sec:fringe_rate_filtering}, the reduced chi-square changes with LST. \autoref{fig:chisqaure_with_filter} shows the reduced chi-square obtained after filtering visibility before calibration with an incomplete sky model. The chi-square has a maximum of 50, even in regions where the diffuse emission is dominant in the $1^{\rm h} < {\rm LST} < 1.8^{\rm h}$ range, compared to 140 obtained without filter, the improvement is at most a factor of 3 in the reduced chi-square. Another way to look at the improvement is that the filter has effectively increased the LST range that can be used for calibration. Even better, the reduced chi-square is smaller than six in the LST range used for calibration, i.e. $[2.2^{\rm h} - 2.8^{\rm h}]$.

\autoref{fig:gains_filtered_data} shows the gains recovered from calibration using an incomplete sky model and after applying a baseline-independent notch filter. Even at the LST range used for calibration, where compact sources dominate, the sky model is still significantly incomplete. Without applying the filter, gains absorb some of the unmodelled diffuse emission on the sidelobes of the primary beam. As a result, gains acquire a pronounced frequency structure, particularly in the $100 < \tau < 500$~ns delay range. The use of a fringe rate baseline-independent notch filter suppresses much of this frequency structure, therefore reducing the excess power seen at delays greater than $\sim 100$~ns. A noticeable result of the filter is that gains now closely match the actual gains up to the noise floor. The most noticeable difference is now a nearly delay-constant offset due to noisier, filtered visibilities. The gains obtained by using different filters have a very similar profile in delay space, indicating that the filter with a 0.25~mHz width is already sufficient to suppress unwanted emission.

The use of a baseline-independent notch filter has a distinct impact on the residual calibrated visibilities in Fourier space.
Let's consider pointing at LST~$ = 1.25^{\rm h}$, where the diffuse emission is the dominant component of the sky emission.  
\autoref{fig:power_spectrum_wedge} shows the effect of the filter in a typical wedge plot in $k_\parallel-k_\perp$, or in the delay spectrum approach equivalently $\tau$ - baseline length space. 
We see that residuals are just noise-like when a complete sky model is used for calibration, as expected. When an incomplete sky model is used, however, the calibrated visibilities differ from the model visibilities due to biased gain calibration solutions. As a result, there is foreground emission in the residual  
confined to the $-500 < \tau < 500$~ns delay range. This contamination is significantly beyond the horizon limit for baselines shorter than $\sim 70$~m.

Residual foreground emission is reduced when calibration is preceded by fringe filtering. The bleed of foreground emission at delays of $|\tau|\sim 300 - 500$~ns is significantly suppressed by more than one order of magnitude. There is no foreground contamination beyond the horizon limit, leaving the EoR window effectively uncontaminated. Foreground residuals remain present at small delays, well within the horizon. Thus even in the worst field, i.e. where the sky model is most incomplete, filtering proves to be very effective at mitigating foreground contamination of the EoR window. This could, in theory, broaden the LST range that could be used for calibration. 

Let's now consider the LST range used for calibration; field centred at LST$= 2^h$. \autoref{fig:power_spectrum_all_filter} shows the residual power spectrum at $z=7.6$. In this field, the foreground contamination can be as high as two orders of magnitude above the noise floor in the $0.15 < \k_{||} < 0.2$~h~Mpc$^{-1}$ range when calibration is carried out with no fringe  rate filter and an incomplete sky model. The behaviour is fairly similar for both the 14 and 28~m baselines considered. Such contamination is concerning, as it appears at $k$ modes where the sensitivity to the EoR is the highest.

\begin{figure*}
    \centering
    \includegraphics[scale=0.55]{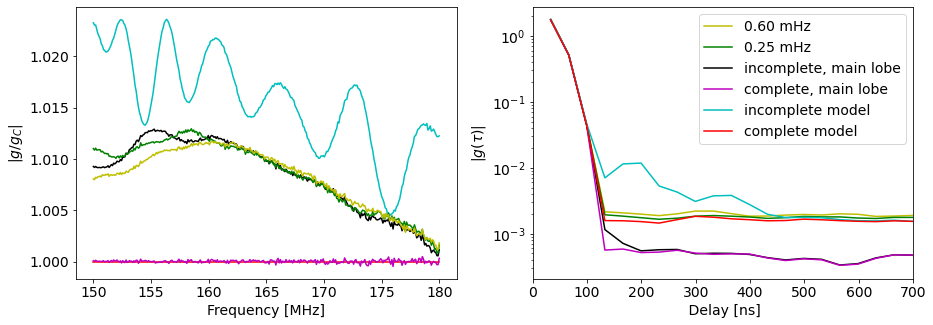}
    \caption{The recovered gains after calibration (similar to \autoref{fig:gains_filtered_data}), now including the application of the baseline-dependent main-lobe filter. Left: The antenna-averaged gain amplitude divided by the average gains estimated with a complete sky model. Right: The antenna-averaged gains in Fourier space. We now show the recovered gains having applied the baseline-dependent main-lobe filter, both to the complete sky model (purple) and an incomplete sky model (black). The baseline-dependent main-lobe filter not only suppresses the spurious structure in the incomplete model (cyan), but also reduces the overall noise floor of the recovered gains, as evidenced by the lower plateau at high delays.}
    \label{fig:gains_Gaussian_filters}
\end{figure*}

\begin{figure*}
    \centering
    \includegraphics[scale=0.55]{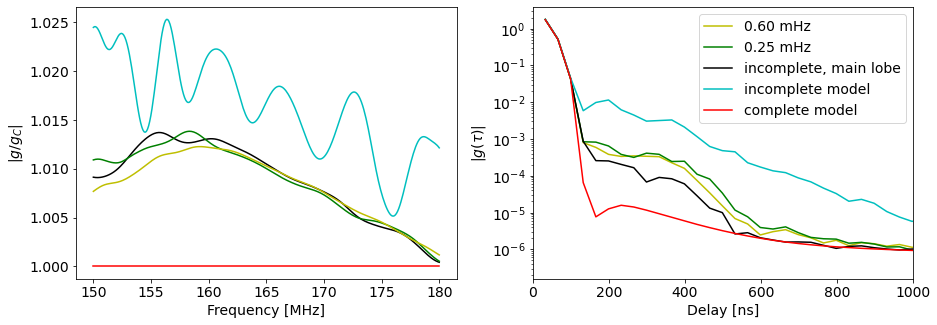}
    \caption{ Same as \autoref{fig:gains_Gaussian_filters} but for a noise-free simulation. This now allows us to probe the performance of the technique down to higher dynamic ranges. The spurious structure in the gains due to model incompleteness (cyan) is suppressed by a over an order of magnitude for $\tau > 150$ ns by the  baseline-dependent main-lobe filter, which also shows slightly better performance than the baseline-independent notch filters for this same delay range. In all, the gains recovered with the baseline-dependent main-lobe filter reach a dynamic range of $\sim10^5$ for $\tau\gtrapprox$ 400 ns.}
    \label{fig:gains_Gaussian_filters_noise_free}
\end{figure*}

The contamination is greatly suppressed if the baseline-independent notch filters are used, leaving the smooth unmodelled emission at $k_\parallel < 0.15\ h\ {\rm Mpc}^{-1}$. Overall, the foreground leakage is suppressed at $k_\parallel \sim 0.2$~h~Mpc$^{-1}$ by at least two orders of magnitude, down to the noise level in 14~m baselines. The filters help suppress unsmooth foreground even on longer baselines, i.e. 28~m baselines, but only up to a factor of ten. This is somewhat expected as the larger baselines are less sensitive to diffuse emission. Notably, the performance of the different filters in suppressing foreground spectral leakage is the same, although evidently, the filters filter varying amounts of foreground emission. Thus, the key to mitigating foreground leakage is to suppress spectrally unsmooth foreground emission, i.e. foreground emission that is coupled with primary beam sidelobes and is also bright enough after primary beam attenuation to drive calibration solution. 

\subsection{Calibration with the main lobe filter}
\label{sec:applying_main_lobe_filters}

We now consider the more aggressive baseline-dependent main-lobe filter. \autoref{fig:gains_Gaussian_filters} shows the gains obtained after calibrating with a filtered complete and incomplete sky model. As expected, the baseline-dependent main-lobe filter attenuates more of the missing foreground emission than the baseline-independent notch filters (\autoref{fig:mainlobe_filter_applied}), resulting in gains that have less spectral structure.
In addition, the baseline-dependent main-lobe filter also suppresses more of the thermal noise in the visibilities than the baseline-independent notch filters, resulting in gain solutions that themselves also have lower noise floors, as evidenced by the lower plateau in \autoref{fig:gains_Gaussian_filters}.
However, as previously noted, the baseline-dependent main-lobe filter correlates noise between different time bins, resulting in gain solutions that are now more correlated between different times.
We defer exploration of these consequences on real HERA analyses to future work.

To understand the performance of these filters to deeper dynamic ranges, we repeated this full analysis on noise-free visibility simulations, shown in \autoref{fig:gains_Gaussian_filters_noise_free}.
Here we can see just how deeply the spurious spectral structure introduced by model incompleteness (cyan) is suppressed.
We see that the baseline-independent notch filters suppress these structures by over an order of magnitude for $\tau>150$ ns, with the baseline-dependent main-lobe filter gaining an additional factor of 2 in suppression, however, recall this comes at the expense of more time-to-time noise correlations.
The choice of which filter is best likely depends on the specific analysis at hand. Nonetheless, we have demonstrated that both are highly effective at suppressing spurious spectral structure in HERA calibration solutions caused by poorly modeled diffuse galactic synchrotron emission.

\subsection{What does this mean for detecting the EoR?}

Having shown how fringe rate filtering the visibilities before calibration can suppress spectral structure in the recovered gains by over an order of magnitude, the question we are left with is, how much is enough?
Is an order of magnitude improvement enough to actually make a 21\,cm EoR detection?
The answer to this question is highly dependent on a number of factors, including the kind of telescope used for observations
(i.e. its primary beam response) and to which part of the sky it is pointed at (i.e. at a hot or cold spot of the foreground sky).
Nonetheless, roughly speaking, it is generally assumed that we need to achieve calibration precision with a dynamic range in the recovered gains of $\sim10^5$ or more to make a fiducial EoR detection for cosmological modes of $k\gtrapprox0.13\ h\ {\rm Mpc}^{-1}$ \citep{DeBoer2017}.
At $z = 8$, this roughly translates to $\tau = 400$~ns.
From \autoref{fig:gains_Gaussian_filters_noise_free}, we can see that, indeed, the recovered gain solutions after fringe rate filtering roughly hit this $10^5$ dynamic range requirement for $\tau>400$~ns, meaning that our technique can (in principle) produce gain solutions with the required precision for a fiducial 21\,cm detection given the assumptions made in this study.
Specifically, this is only a statement on our technique's ability to mitigate spectral structure due to unmodeled diffuse foregrounds, not on its ability to mitigate other real-world factors, such as crosstalk and/or mutual coupling, which we defer to future work.

%%%%%%%%%%%%%%%%%%%%%%%%%%%%%%%%%%%%%%%%%%
%%%%%%%%%%%%%%%%% Conclusion %%%%%%%%%%%%%
%%%%%%%%%%%%%%%%%%%%%%%%%%%%%%%%%%%%%%%%%%
\section{Conclusions}
\label{sec:conclusions}

In this paper, we have presented a technique for mitigating the impact of poorly modelled Galactic diffuse foregrounds in the calibration of 21~cm drift-scan observations.The technique relies upon the use of temporal filters that isolate and suppress diffuse emission in the sidelobes of the primary beam, while retaining the well-understood point source emission in the main lobe of the primary beam.
We explored two different types of filters: a baseline-independent ``notch'' filter, and a baseline-dependent ``main lobe'' filter.
Notch filters suppress only foreground emission centred at the zero fringe rate mode in the visibilities, whereas the baseline-dependent main-lobe filters aim to suppress all emission that is inconsistent with point sources in the field of view of the primary beam.

To test our technique, we simulated realistic HERA observations where we included point sources, extended sources, and a diffuse galactic component in our sky model \citep{Hurley-Walker2017, Zheng2016}, and used an electromagnetic simulation of the HERA primary beam response \citep{Fagnoni2019}.

The simulated visibilities were corrupted with realistic mock gains and then pushed through HERA's redundant and absolute calibration pipeline.
We calibrate against a shallow point source catalogue in order to simulate a realistic observation where diffuse emission is not considered and the point source catalogue is incomplete.
We test whether our visibility filters improve the quality of the subsequent recovered gain solutions, relative to the case where no filters are used.
We found biased calibration solutions when no filters are used, even if we adopt the standard procedure of discarding short baselines that are most sensitive to diffuse emission.
This bias takes the form of spurious frequency structure in the gains that appears at $\tau \gtrsim 150$~ns in the gains in Fourier space.
This directly impacts the 21\,cm power spectra, resulting in increased foreground leakage for $k>0.15$ h Mpc$^{-1}$ when these calibration solutions are applied to the data.

The use of fringe rate filters before calibration substantially mitigates these biases, reducing the excess spectral structure in the recovered gain solutions by over an order of magnitude.
This reduces the observed foreground leakage down to the simulated noise floor in the data at $k>0.15$ h Mpc$^{-1}$.
When repeating this analysis on noiseless simulations, we find that our technique yields gain solutions that are accurate at 1 part in $10^5$ in dynamic range for delays $\tau\gtrsim400$ ns, which is roughly at the estimated calibration requirement for detecting the EoR with HERA at $k\gtrsim0.2$ h Mpc$^{-1}$ \citep{DeBoer2017}.

This technique can be directly applied to improve the calibration of drift scan observation from radio interferometric arrays, such as the Canadian Hydrogen Intensity Mapping \citep[CHIME,][]{CHIMEcollaboration2022} aimed at mapping neutral hydrogen over the redshift range $z=0.8-2.5$. However, for other interferometric arrays probing the EoR, which are mainly observing in tracking mode, such as Square Kilometer Array \citep[SKA,][]{Koopmans2015} and Low-Frequency Array \citep[LOFAR,][]{VanHaarlem2013}, our technique be cannot be applied.

Future work will explore applying these fringe rate filters to the problem of mitigating non-redundancies, for example, due to primary beam variations across the array \citep{Orosz2019}, and mutual coupling \citep{Fagnoni2019, Kern2019, Kern2020b, Josaitis2022}, as well as looking at ways of implementing these filters in a way that is robust to radio frequency interference.

\section*{Acknowledgements}

The authors gratefully acknowledge helpful discussions with Joshua Dillon, Jacqueline Hewitt, Jonathan Pober, Aaron Parsons, Bryna Hazelton and Miguel Morales. N.C. acknowledges the financial assistance of the South African Radio Astronomy Observatory (SARAO, \href{www.sarao.ac.za}{www.sarao.ac.za}). N.K. acknowledges support from the MIT Pappalardo fellowship. N.F and E.L acknowledge the financial support of the UKRI STFC (United Kingdom Research and Innovation Science Facility Council).

\section*{Data Availability}
Data used in this work may be made available upon reasonable request to the corresponding author. The code used for the analysis of this work is publicly available at \href{https://github.com/Ntsikelelo-Charles/Fringe_rate_filters}{https://github.com/Ntsikelelo-Charles/Fringe\_rate\_filters}. This work relied on publicly available and open-sourced community Python software, including \texttt{numpy} \citep{2020NumPy-Array}.



\bibliographystyle{mnras}
\bibliography{eor_references} % if your bibtex file is called example.bib

%%%%%%%%%%%%%%%%%%%%%%%%%%%%%%%%%%%%%%%%%%%%%%%%%%

%%%%%%%%%%%%%%%%% APPENDICES %%%%%%%%%%%%%%%%%%%%%

%%%%%%%%%%%%%%%%%%%%%%%%%%%%%%%%%%%%%%%%%%%%%%%%%%

% Don't change these lines
\bsp	% typesetting comment
\label{lastpage}
\end{document}